\documentclass[aps,prl,twocolumn,nofootinbib,superscriptaddress,preprintnumbers,longbibliography]{revtex4-2}
\usepackage[utf8]{inputenc} 
\usepackage[T1]{fontenc}
\usepackage{ucs}
\usepackage{epsfig}
\usepackage{graphicx}
\usepackage[english]{babel}
\usepackage{hyphenat}
\usepackage{amsmath}
\usepackage{amssymb}
\usepackage{bbold}
\usepackage{mathtools}
\usepackage{mathrsfs}
\usepackage{slashed}
\usepackage{epstopdf}
\usepackage[dvipsnames]{xcolor}
\usepackage{braket}
\usepackage{booktabs}
\definecolor{lcolor}{rgb}{0.,0.0,0.}
\definecolor{citcolor}{rgb}{0,0.,0.5}
\usepackage[breaklinks,colorlinks,urlcolor=blue,citecolor=blue,linkcolor=blue]{hyperref}
\usepackage{multirow}
\usepackage{ltablex}
\usepackage{soul}
\usepackage{feynmp-auto}
\usepackage{tikz}
\usetikzlibrary{decorations.pathmorphing,decorations.markings,arrows.meta,positioning,calc}

\usepackage[e]{esvect}

\usepackage{media9}

\def\cV{\mathcal{V}}

\def\cG{{\cal G}}

\def\cP{{\cal P}}
\def\cK{{\cal K}}
\def\cM{{\cal M}}
\def\cD{{\cal D}}

\def\cN{{\cal N}}
\def\cR{{\cal R}}

\def\cW{{\cal W}}

\def\beps{{\boldsymbol \epsilon}}

\newcommand{\secn}[1]{Section~1}
\newcommand{\appn}[1]{Appendix~1}

\long\def\comment#1{ }

\def\and{\quad\text{and}\quad}

\def\q{{\boldsymbol q}}
\def\0{{\boldsymbol 0}}
\def\1{{\boldsymbol 1}}
\def\p{{\boldsymbol p}}
\def\l{{\boldsymbol l}}
\def\k{{\boldsymbol k}}
\def\x{{\boldsymbol x}}
\def\y{{\boldsymbol y}}

\def\r{{\boldsymbol r}}

\def\u{{\boldsymbol u}}

\def\0{{\boldsymbol 0}}

\renewcommand{\l}{\boldsymbol{l}}

\renewcommand\a{\alpha}
\renewcommand\d{\delta}

\renewcommand\t{\tau}
\renewcommand\u{\upsilon}

\renewcommand\o{\omega}
\newcommand\e{\epsilon}
\newcommand\m{\mu}

\renewcommand\O{\Omega}

\newcommand{\re}{{\rm{Re}}}

\def\u{{\boldsymbol u}}

\renewcommand{\part}{{\rm part}}

\newcommand{\be}{\begin{equation}}
\newcommand{\ee}{\end{equation}}
\newcommand{\bes}{\begin{subequations}}
\newcommand{\ees}{\end{subequations}}
\newcommand{\bea}{\begin{eqnarray}}
\newcommand{\eea}{\end{eqnarray}}

\newcommand{\nn}{\nonumber \\}
\newcommand{\na}{\nabla}

\def\bea#1\eea{\begin{align}#1\end{align}}
\newcommand{\bef}{\begin{figure}[h!tb]\centering}
\newcommand{\eef}{\end{figure}}

\allowdisplaybreaks

\begin{document}

\title{Directional dead-cone effect in QCD matter}

\author{Jo\~{a}o Barata}
\email{joao.lourenco.henriques.barata@cern.ch}
\affiliation{CERN, Theoretical Physics Department, CH-1211, Geneva 23, Switzerland}

\author{Matvey V. Kuzmin}
\email[]{matvei.v.kuzmin@gmail.com}
\affiliation {Instituto Galego de F{\'{i}}sica de Altas Enerx{\'{i}}as,  Universidade de Santiago de Compostela, Santiago de Compostela 15782,  Spain}
\author{Xo{\'{a}}n Mayo L\'{o}pez}
\email[]{xoanml@mit.edu}
\affiliation{Center for Theoretical Physics – a Leinweber Institute, Massachusetts Institute of Technology,
Cambridge, MA 02139}
\affiliation{Instituto Galego de F{\'{i}}sica de Altas Enerx{\'{i}}as,  Universidade de Santiago de Compostela, Santiago de Compostela 15782,  Spain}

\author{Andrey V. Sadofyev}
\email{andrey.sadofyev@ehu.eus}
\affiliation{Department of Physics, University of the Basque Country UPV/EHU, P.O. Box 644, 48080 Bilbao, Spain}
\affiliation{IKERBASQUE, Basque Foundation for Science, Plaza Euskadi 5, 48009 Bilbao, Spain}
\affiliation{Laboratório de Instrumentação e Física Experimental de Partículas (LIP), Av. Prof. Gama Pinto, 2, 1649-003 Lisbon, Portugal}

\author{Carlos A. Salgado}
\email[]{carlos.salgado@usc.es}
\affiliation{Instituto Galego de F{\'{i}}sica de Altas Enerx{\'{i}}as,  Universidade de Santiago de Compostela, Santiago de Compostela 15782,  Spain}

\preprint{CERN-TH-2025-237} 
\preprint{MIT-CTP/5964}

\begin{abstract}
We consider the propagation of heavy quarks through a dense, hydrodynamically flowing QCD medium, representative of the quark–gluon plasma formed in ultrarelativistic heavy-ion collisions. Working in the high-energy limit, we identify two novel mass-dependent effects arising from the heavy quark coupling to the local medium flow. The first is the emergence of a tensorial jet transport coefficient, $\hat q_{ij}$, which encodes the directional structure of transverse-momentum broadening. The second, named the \textit{directional dead-cone} effect, corresponds to an anisotropic suppression of medium-induced radiation aligned with the hydrodynamic flow. We discuss how these effects manifest in jet observables and identify distinctive signatures of heavy quark dynamics in an evolving medium.
\end{abstract}

\maketitle

\noindent\emph{\textbf{Introduction:}} A central question in ultrarelativistic heavy-ion collisions (HICs) is how the medium modifies the evolution of energetic partons produced in the same event, see e.g.~\cite{Apolinario:2022vzg} for a recent review. Such medium-induced modifications provide a powerful tomographic probe of the space–time structure of the quark–gluon plasma (QGP) and offer direct insight into the properties of QCD matter under extreme conditions, complementary to soft-sector observables. Advancing our theoretical understanding of how hard probes evolve within the medium, and how this manifests in experimental observables, is therefore essential for a quantitative characterization of the matter produced in HICs.

In this context, energetic partons---which subsequently form collimated sprays of particles known as jets---traverse the QGP and undergo multiple scatterings that induce bremsstrahlung, transferring energy from the jet to the medium. Although the structure of medium-induced radiation has been extensively studied \cite{Apolinario:2022vzg}, it remains challenging to access experimentally. Heavy quarks offer a particularly clean avenue toward this goal: their finite mass suppresses radiation within the characteristic angular region set by the dead-cone angle $\theta_{\rm dc}\equiv m_Q/E$, with $m_Q$ the heavy quark mass and $E\equiv p^+$ the total energy of the system, limiting virtuality-driven emissions around the quark direction~\cite{Dokshitzer:1991fd}. Medium-induced radiation is not subject to this kinematic suppression, leading to the filling of the dead-cone through secondary emissions~\cite{Armesto:2003jh}, see also~\cite{Dokshitzer:2001zm,Djordjevic:2003zk,Zhang:2003wk,Aurenche:2009dj}. Experimental access to the dead-cone angle is thus particularly valuable; it has only recently been observed in proton–proton collisions~\cite{ALICE:2021aqk,Cunqueiro:2018jbh}, and some recent works~\cite{Cunqueiro:2022svx,Li:2017wwc} aim to expand such measurements to HIC environments.

\begin{figure}[h!]
    \centering
    \includegraphics[width=0.7\linewidth]{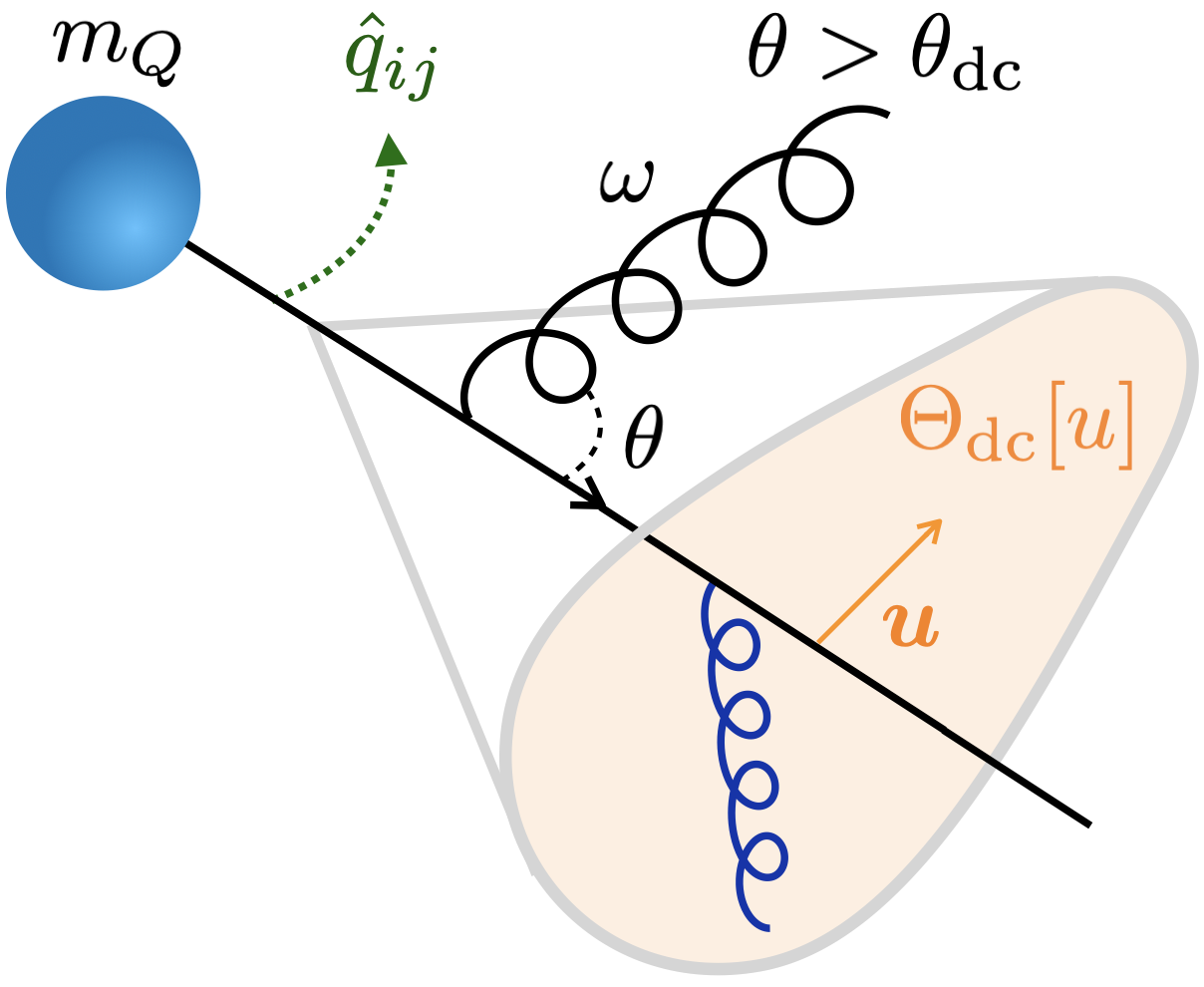}
    \caption{Heavy quark propagation in a flowing medium with velocity $u^\mu$. Momentum broadening takes a tensorial character, introduced via $\hat q_{ij}$; homogeneous diffusion is recovered in the massless quark limit. A \textit{directional dead-cone} angle $\Theta_{\rm dc}$ leads to a modulation of the dead-cone geometry and medium induced radiation pattern inside.}
    \label{fig:dead-cone_cartoon}
\end{figure}

Previous works on high-energy heavy quarks in HICs have mainly focused on the longitudinal dynamics, with respect to the leading parton momentum, governed by the interplay of the dead-cone effect, and the Landau–Pomeranchuk–Migdal (LPM) QCD effect~\cite{Landau:1953um,Migdal:1956tc, Baier:1996sk, Baier:1996kr, Zakharov:1996fv, Zakharov:1997uu, Gyulassy:2000fs, Gyulassy:2000er,Wang:2001ifa,Arnold:2002ja}, whose characteristic angle, $\theta_{\rm c}\equiv 1/\sqrt{\hat{q}_0L^3}$, with $\hat{q}_0$ the jet-quenching parameter and $L\equiv L^+$ the light-cone extent of the medium, sets the angular region below which medium-induced \textit{bremsstrahlung} is coherently enhanced. Thus, gluon radiation at angles $\theta < \theta_{\rm c} < \theta_{\rm dc}$ is parametrically dominated by medium-induced emissions, while virtuality-driven fragmentation is power suppressed.

In this letter, we argue that the dead-cone region can serve as a more powerful diagnostic tool of the bulk, as it can probe its spatial (transverse) dynamics, via the unique connection between heavy quarks and the hydrodynamic 
gradients and the matter's flow. The coupling between the heavy quark and the flowing medium induces characteristic deformations of the radiation pattern and the dead-cone geometry, exhibiting sensitivity to the spatial evolution of the QGP. We identify two novel effects absent for light partons: \textbf{(i)} the emergence of an effective tensorial jet transport coefficient, $\hat q_{ij}$, which reduces to the conventional scalar $\hat q_0$ in the massless limit; and \textbf{(ii)} a directional dead-cone effect, corresponding to a stronger suppression of radiation emitted along the direction of the medium's flow. We further discuss how these features manifest in jet observables, establishing a clear path toward their phenomenological implementation in HICs. These ideas are succinctly summarized in Fig.~\ref{fig:dead-cone_cartoon}. 

\vspace{0.25cm}
\noindent\emph{\textbf{Tensorial momentum diffusion:}} We consider first the evolution of a single heavy quark in the presence of a medium 
flowing ultrarelativistically with a light-like flow field $u^\mu=\left(u^+,u^-,\u\right)^\mu=\left(\frac{1-\sqrt{1-\u^2}}{\sqrt{2}}, \frac{1+\sqrt{1-\u^2}}{\sqrt{2}},\u\right)^\mu$.\footnote{Notice that the medium flow model differs from the one used in the earlier works on the jet-flow interactions~\cite{Sadofyev:2021ohn,Andres:2022ndd,Kuzmin:2023hko,Kuzmin:2024smy}. This choice considerably simplifies the discussion of the medium-induced radiation and its gauge invariance.} The medium is modeled as a stochastic color field, incoherently generated by its flowing quasi-particles:
\begin{align}\label{eq: field def}
\notag gA^{a\mu}(q)&=(2\pi)\, v(-q^2)\, u^\mu\, \delta(q \cdot u )
\\
&\times\int_{y^+,\y} \rho^a(\y,y^+)e^{iq^- y^+} e^{-i\q \cdot \y}\,.
\end{align}
Here $\rho^a(y^+,\y)$ is the color density of the sources in the bulk, whose interaction with the heavy quark is determined by a model-dependent scattering potential $v(-q^2)$, accounting for medium Debye screening. 

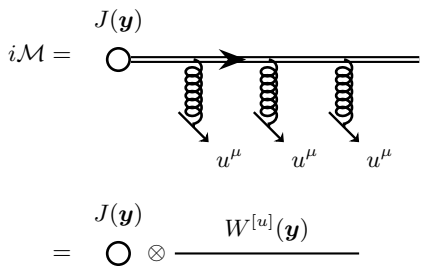
\begin{figure}[h!]
  \centering
  \[
\begin{aligned}
i \mathcal{M}
&=\;
\begin{tikzpicture}[
    baseline=-0.1ex,
    >=Stealth
]
  \tikzset{
   heavy/.style={
  double,
  line width=0.8pt,
  double distance=1pt,
  postaction={decorate},
  decoration={markings,
    mark=at position 0.4 with {\arrow[scale=.7]{>}} 
  }
},
    uvec/.style={
      line width=0.9pt,
      -{Stealth[length=3pt,width=5pt]}
    }
  }

  \node[draw,circle,inner sep=3pt,line width=1pt] (J1) {};
  \node[above=2pt of J1] {$J(\y)$};

  \coordinate (end1) at (4,0);
  \draw[heavy] (J1) -- (end1);

  \foreach \x in {1.0,2.0,3.0} {
    \coordinate (gend) at (\x,-0.9);
    \draw[decorate,
          decoration={coil,aspect=0.8,segment length=3pt,amplitude=3pt},
          line width=1pt]
          (\x,0) -- (gend);
    \draw[uvec] (\x-0.2,-0.7) -- (\x+0.2,-1.1)
   node[below right=-1pt] {$u^\mu$};
  }
\end{tikzpicture}
\\[0.4em]
&=\;
\begin{tikzpicture}[
    baseline=-0.5ex
]
  \node[draw,circle,inner sep=3pt,line width=1pt] (J2) {};
  \node[above=2pt of J2] {$J(\y)$};

  \node[right=0.1cm of J2] (X) {$\otimes$};

  \coordinate (end2) at (3.2,0);
\draw[line width=0.8pt] (X) -- node[above=1pt] {$W^{[u]}(\y)$} (end2);
\end{tikzpicture}
\end{aligned}
\]
\caption{Amplitude for the evolution of a single heavy quark in the presence of a flowing medium with velocity $u^\mu$. The lower line indicates that, to the accuracy considered here, the amplitude can be cast as a convolution of the initial quark current $J$ and a time-like Wilson line $W$, now dependent on the background flow field.}
\label{fig:diagram_broad}
\end{figure}

The evolution of a heavy quark in the medium, within a perturbative framework, is captured by amplitudes containing an arbitrary number of tree-level gluon exchanges with the bulk, as illustrated in Fig.~\ref{fig:diagram_broad}. The amplitude with $N$ gluon exchanges takes the form
\begin{align}\label{eq:amp_B_1}
i\mathcal{M}_N&=\int\Bigg[\prod^N_{n=1}\frac{d^2\p_ndp^-_n}{(2\pi)^3}\,(-1)\,t^{a_n}\,\rho^{a_n}(p_{n+1}-p_n)\nn
&\hspace{0cm}\times v\left(-(p_{n+1}-p_n)^2\right)\frac{2\, p\cdot u}{u^-(p_n^2-m_Q^2)}\Bigg]J(p_1)\,,
\end{align}
where $J(p)$ is the source generating the initial heavy quark, and $p = p_{N+1}$ is its final momentum. This amplitude simplifies significantly in the high-energy limit, where the final-state quark energy is taken to be larger than the characteristic transverse scale $\bot$. In this regime, only the poles of the quark propagators contribute to the $p_n^-$ integrals, while the residues corresponding to the poles of $v\left(-(p_{n+1}-p_n)^2\right)$ are exponentially suppressed due to the Debye screening, and can be neglected in a sufficiently dilute and longitudinally-extended medium. We further focus on the leading-power mass effects, assuming the ordering $|\p|/E \ll m^2_{\rm Q}/E^2 <1 $, such that certain kinematic corrections can be neglected in comparison to the mass-dependent ones. Summing over all the possible numbers of gluon exchanges with the medium, the resummed amplitude reduces to
\begin{align}\label{eq:amp_B_2}
i\mathcal{M}&=\sum_{N=0}^\infty i \mathcal{M}_N=\int_{\y}\,e^{-i\p\cdot\y} J(\y) \nn 
&\times \mathcal{P}\text{exp}\left\{i\int_0^L d\tau\,t^{a}v^a\left(\y-\frac{1}{2} \frac{m_Q^2}{E^2}\frac{\u}{u^-}\tau, \tau\right)\right\} \nn 
&= J(\y) \, \otimes W^{[u]}(\y;L,0) \,.
\end{align}
In Eq.~\eqref{eq:amp_B_2} the flow enters through modifications of the vertices and propagator poles, see details in Appendix and~\cite{Andres:2022ndd} for the discussion for massless quarks. The last line in Eq.~\eqref{eq:amp_B_2} highlights that, within the theoretical accuracy of this calculation, the amplitude for the evolution of the heavy quark can still be written as the convolution of an initial current with a path-ordered exponential along the light-like path of the quark, as typically found in the related literature for static matter~\cite{Casalderrey-Solana:2007knd}. The flow effects can be recast as a modification of the color scattering potential $v^a$:
\begin{align}\label{eq:vamodified}
v^a(\x,x^+)&=\left(1+\frac{u^+}{u^-}\frac{m_Q^2}{E^2}\right)\int_{\q}\rho^a(\q,x^+)\nn
&\times v\left(\q^2+\frac{m_Q^2}{E^2}\left[\frac{\u\cdot\q}{u^-}\right]^2\right)e^{i\q\cdot \x}\,.
\end{align}

Upon squaring Eq.~\eqref{eq:amp_B_2} and averaging the cross-section  over the color sources $\rho^a$, one can directly compute the effects of momentum diffusion in the evolution of the heavy quark in flowing media. Assuming Gaussian statistics with constant number density $\rho$,
$2 C_A\langle\rho^a(\x,x^+)\rho^b(\bar{\x},\bar{x}^+)\rangle
=\rho \, \delta^{ab}\delta^{(2)}(\x-\bar{\x})\delta(x^+-\bar{x}^+)
$
the process is determined by the following two-point function of Wilson lines:
\begin{align}\label{eq:amp_B_WW}
    &\log \langle W^{[u]}(\y;L,0)W^{[u]\dag}(\bar \y;L,0)\rangle   = \frac{C_F}{2N_c} L\rho \left(1+\frac{u^+}{u^-}\frac{m_Q^2}{E^2}\right)^2 \nn 
    &\times \int_{\q} \left(e^{i\q\cdot(\y-\bar{\y})} - 1\right) v^2\left(\q^2+\frac{m_Q^2}{E^2}\left[\frac{\u\cdot\q}{u^-}\right]^2\right) \, ,
\end{align}
where the flow-velocity argument shifts introduced for the Wilson line in Eq.~\eqref{eq:amp_B_2} vanish due to the time locality of the source-averaged interactions. We can directly extract one of the most phenomenologically relevant quantities, the jet transport coefficient, from Eq.~\eqref{eq:amp_B_WW}. Its straightforward extension to anisotropic momentum diffusion, see e.g. \cite{Romatschke:2006bb,Dumitru:2007rp,Baier:2008js}, reads:
\begin{align}\label{eq:qhat_flowing}
\hat{q}_{ij} & = - \frac{\partial}{\partial L}\big[\na_{(y-\bar{y})_i}\na_{(y-\bar{y})_j} \nn
& \hspace{1cm} \times\langle W^{[u]}(\y;L,0)W^{[u]\dag}(\bar \y;L,0)\rangle\big]_{\bar{\y}=\y}\nn 
&=\frac{\hat{q}_0}{2}\left(1+\frac{u^+}{u^-}\frac{m_Q^2}{E^2}\right)\left[\delta_{ij}-\frac{u_iu_j}{(u^-)^2}\frac{m_Q^2}{E^2}\right]\,,
\end{align}
where a specific model potential $v(\q^2)=-\frac{g^2}{\q^2+\mu^2}$, with screening mass $\mu$ and strong coupling constant $g$, has been used \cite{Gyulassy:1993hr}. The scalar jet quenching parameter is given by its trace, $\hat q \equiv \sum_i \hat q_{ii}$, and $\hat{q}_0$ can be identified with the transport coefficient in the massless quark limit. The physical interpretation of Eq.~\eqref{eq:qhat_flowing} is immediate: as an ensemble of heavy quarks propagates through the medium, it is distorted by the flow, which not only changes the local density but also transfers transverse momentum to the quarks. This induces anisotropy in the transverse momentum diffusion of the ensemble, making it direction-dependent within the transverse plane. Specifically, the interaction between the heavy-quark mass and the medium flow leads to different momentum-broadening strengths parallel and orthogonal to the flow direction. One should also note that this anisotropy pattern differs from the averaged transverse momentum acquired by light quarks through momentum broadening in evolving matter, see~\cite{Sadofyev:2021ohn,Antiporda:2021hpk,Barata:2022krd,Andres:2022ndd,Barata:2022utc,Kuzmin:2023hko,Bahder:2024jpa}, and is instead more similar to the tensorial anisotropy arising from higher-order sub-eikonal corrections to light-quark dynamics, see e.g. \cite{Barata:2022utc,Barata:2025htx}, or to that discussed in the context of probe-matter interactions during the initial stages of HICs, see~\cite{Carrington:2021dvw,Hauksson:2021okc,Avramescu:2023qvv,Hauksson:2023tze,Boguslavski:2024ezg,Boguslavski:2024jwr,Barata:2024bqp,Silva:2025dan}.

\vspace{0.25cm}
\noindent\emph{\textbf{Directional dead-cone effect:}} The mass-flow coupling effects also play a role in the production of radiation off the heavy quark, as indicated in Fig.~\ref{fig:dead-cone_cartoon}. The amplitude for this process can again be written in terms of an arbitrary number of insertions of soft-gluon exchanges with the flowing medium. The calculation of this process at full sub-eikonal accuracy is rather intricate; here we instead consider the soft gluon limit, in which the gluon light-cone energy $\o$ is taken to be small compared to the total energy of the daughter partons in the final state $\omega\equiv x E\ll E$. We keep only the terms enhanced by powers of $x^{-1}$ or by $m_Q$, as well as those scaling with the length up to $\mathcal{O}\left(\left(\frac{\bot^2}{x E } \, x^+\right)^n\right)$ and $\mathcal{O}\left(\left(\frac{m^2_{Q}}{E^2}  \, \bot \, x^+\right)^n\right)$, \textit{i.e.} the LPM phases and their leading mass corrections.\footnote{Notice that the anisotropic nature of $\hat{q}_{ij}$ goes beyond this approximation. In a more general case, this effect would also be present, making the calculation more involved, and therefore masking the simple interpretation of the results presented in what follows.}

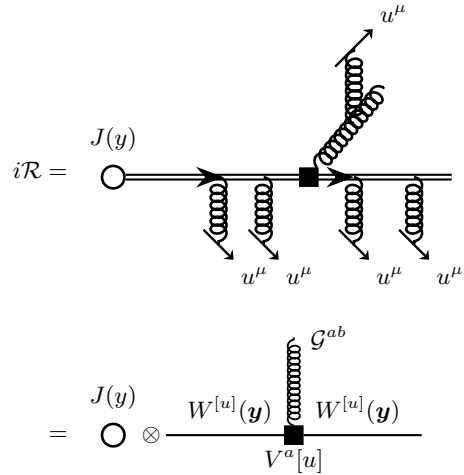
\begin{figure}[h!]
  \centering
  \[
\begin{aligned}
i\mathcal{R}
&=\;
\begin{tikzpicture}[baseline=-0.1ex,>=Stealth]
  \tikzset{
    heavy/.style={
      double,
      line width=0.8pt,
      double distance=1pt,
      postaction={decorate},
      decoration={markings,
        mark=at position 0.3 with {\arrow[scale=.7]{>}},
        mark=at position 0.7 with {\arrow[scale=.7]{>}}
      }
    },
    gluonTop/.style={
      decorate,
      decoration={coil,aspect=0.8,segment length=3pt,amplitude=3pt},
      line width=1pt
    },
    uvec/.style={
      line width=0.9pt,
      -{Stealth[length=3pt,width=5pt]}
    }
  }

  \node[draw,circle,inner sep=3pt,line width=1pt] (J1) {};
  \node[above=2pt of J1] {$J(y)$};

  \coordinate (end1) at (4.5,0);
  \draw[heavy] (J1) -- (end1);

  \node[draw,fill=black,minimum width=7pt,minimum height=7pt] (Vh) at (2.6,0) {};

  \coordinate (Gtop) at (3.6,1.2);
  \draw[gluonTop] (Vh) -- (Gtop);

  \foreach \x in {1.4,2.0,3.2,4.0} {
    \draw[gluonTop] (\x,0) -- +(0,-0.9);
    \draw[uvec] (\x-0.2,-0.7) -- (\x+0.2,-1.1)
      node[below right=-1pt] {$u^\mu$};
  }

  \coordinate (gattach) at ($(Vh)!0.6!(Gtop)$);
  \coordinate (gend)    at ($(gattach)+(0,1.0)$);
  \draw[gluonTop] (gattach) -- (gend);
  \draw[uvec] ($(gend)+(-0.25,-0.25)$) -- ($(gend)+(0.25,0.25)$)
    node[above right=-1pt] {$u^\mu$};

\end{tikzpicture}
\\[0.6em]
&=\;
\begin{tikzpicture}[baseline=-0.5ex,>=Stealth]
  \tikzset{
    gluonBottom/.style={
      decorate,
      decoration={coil,aspect=.8,segment length=2.3pt,amplitude=2.3pt},
      line width=0.7pt
    }
  }

  \node[draw,circle,inner sep=3pt,line width=1pt] (J2) {};
  \node[above=2pt of J2] {$J(y)$};

  \node[right=0.1cm of J2] (ot) {$\otimes$};

  \coordinate (WLstart) at ($(ot.east)+(-0.05,0)$);
  \coordinate (WLend)   at ($(WLstart)+(3.4,0)$);

  \node[draw,fill=black,minimum width=7pt,minimum height=7pt] (V) at ($(WLstart)!0.5!(WLend)$) {};

  \draw[line width=0.8pt] (WLstart) -- (V);
  \draw[line width=0.8pt] (V) -- (WLend);

  \node[above=1pt] at ($(WLstart)!0.5!(V)$) {$W^{[u]}(\y)$};
  \node[above=1pt] at ($(V)!0.5!(WLend)$) {$W^{[u]}(\y)$};
  \node[below=2pt] at (V) {$V^a[u]$};

  \coordinate (G2) at ($(V)+(0,1.3)$);
  \draw[gluonBottom] (V) -- (G2);
  \node[right=3pt] at (G2) {$\mathcal{G}^{ab}$};
\end{tikzpicture}
\end{aligned}
  \]
  \caption{Amplitude entering inclusive single gluon production in the presence of a flowing medium with velocity $u^\mu$. Bottom diagram depicts the decomposition of the amplitude at the leading sub-eikonal order, where all flow effects can be absorbed into the vertex and fermionic lines.}
  \label{fig:diagram_M_final}
\end{figure}

With these approximations, the quark lines can be treated as for the case of heavy quark momentum broadening, and only the gluon line requires further simplification. Working in the axial gauge with $u\cdot A\propto u^2 = 0$, we write the gluon propagator as
\begin{align}\label{eq: gluon prop}
   N_{\mu\nu} \equiv i(k^2+i \e) \, G_{\mu \nu}(k) = g_{\mu\nu}-\frac{k_\mu u_\nu + k_\nu u_\mu}{(k\cdot u)} \, .
\end{align}
The polarization vector satisfies $u\cdot\e^*(k)=0$, while the residual gauge freedom is fixed by imposing $k\cdot\e^*(k)=0$, restricting to the two physical gluon polarizations:
\begin{align}\label{eq: pol vec}
    \e^{*\m}(k) = \left(\beps \cdot \left[ \frac{\u}{u^-} - \k \frac{u^+}{u^-\o}\right], \frac{\beps\cdot \k}{\o},\beps\right)^\mu \, .
\end{align}
Since $N_{\mu\nu}$ and $\e^{*\m}(k)$ are transverse to $u_\mu$, all the three-gluon vertices can be significantly simplified leading to the following form of the resummed emission amplitude
\begin{widetext}
 \begin{align}\label{eq: resummed amplitude not simplified_text}
     i\cR \simeq &  \frac{ig}{\o} \lim_{x^+_f \rightarrow \infty} \int_{x_s^+,\k_1,\x}  J(\x) \,  e^{-i \x\cdot(\l+\k_1)} \, e^{i x^+_s \left(\frac{m_Q^2 x^2}{2\o}-\frac{m_Q^2}{E^2}\frac{(\k-\k_1)\cdot\u}{2u^-}\right)}  \, \nn
     & \times  \beps \cdot
    \k_1 \, W^{[u]}(\y;\infty,x^+_s) \, t^{b_1} \, W^{[u]}(\x;x^+_s,0) \, e^{ix_f^+ \frac{\k^2}{2\o}} \, \cG^{b_f b_{1}}(\k,x^+_f;\k_1,x^+_s)  \,,
\end{align}   
\end{widetext}
where $\l$ and $\k$ denote the final transverse momenta of the quark and gluon, respectively. At the current working accuracy, the color scattering potential $v^a$ inside the Wilson lines reduces to the massless limit form, while the associated $u^\mu$ correction enters only through its argument shift. 
The decomposition of the amplitude into effective scalar propagators and a vertex function is shown at the bottom of Fig.~\ref{fig:diagram_M_final}. The resummation of multiple interactions of the gluon with the matter reduces to the form found in the massless quark case, and is encoded in the single-particle propagator $\cG$.
This is expected as the gluon is massless and receives no velocity corrections at leading eikonal order \cite{Sadofyev:2021ohn,Andres:2022ndd,Kuzmin:2023hko},
see the Appendix for further discussion.

One can directly compute the radiative spectrum from the emission amplitude in Eq.~\eqref{eq: resummed amplitude not simplified_text}; summing its square over final quantum numbers and averaging over the initial ones, as well as averaging over the background field configurations, we find 
\begin{widetext}
   \begin{align} \label{eq:spec_final}
    (2\pi)^2 \o \frac{dI}{d\o d^2\k} &=   \lim_{x^+_f\rightarrow\infty} \frac{\alpha_s}{N_c \, \o^2} \re \int_0^\infty d\bar{x}^+_s \int_0^{\bar{x}^+_s} dx^+_s \int_{\y} \, e^{-i \frac{m_Q^2 x^2}{2\o}  \left(1-\frac{\k\cdot\u}{u^- \o} - \frac{m_Q^2}{E^2} \frac{\u^2}{4 u^{-2}}\right) \left(\bar{x}^+_s - x^+_s \right)} \, \notag
    \\ & \hspace{0cm} \times \, e^{-i \frac{m_Q^2}{E^2} \o \frac{\y\cdot\u}{2u^-}}   \, S_2(\k,\k,x^+_f;\y,\r,\bar{x}^+_s) \, \boldsymbol{\na}_{\y} \cdot  \boldsymbol{\na}_{\x}\,   \cK(\y,\bar{x}^+_s; \x,x^+_s) \Big|_{\r=\x = \boldsymbol{0}}   \, .
\end{align} 
\end{widetext}
The radiation spectrum in Eq.~\eqref{eq:spec_final} is fully determined by two color-singlet structures, $S_2$ and $\cK$, which can be expressed as correlators of the effective propagators discussed above. The correlator $\cK$ governs the production of the emitted soft gluon, while $S_2$ describes the final state momentum broadening of the produced gluon; their form is discussed in the Appendix. For numerical evaluation, we further employ the harmonic oscillator approximation, $\frac{2C_F}{N_c}L\rho\int_{\q} \left(1 - e^{i\q\cdot\r}\right) v^2\left(\q^2\right) \simeq \hat{q}_0\r^2$, 
reducing to the potential in the massless case at the current accuracy.
\begin{figure*}[t!]
    \centering
    \includegraphics[height=0.19\textheight]{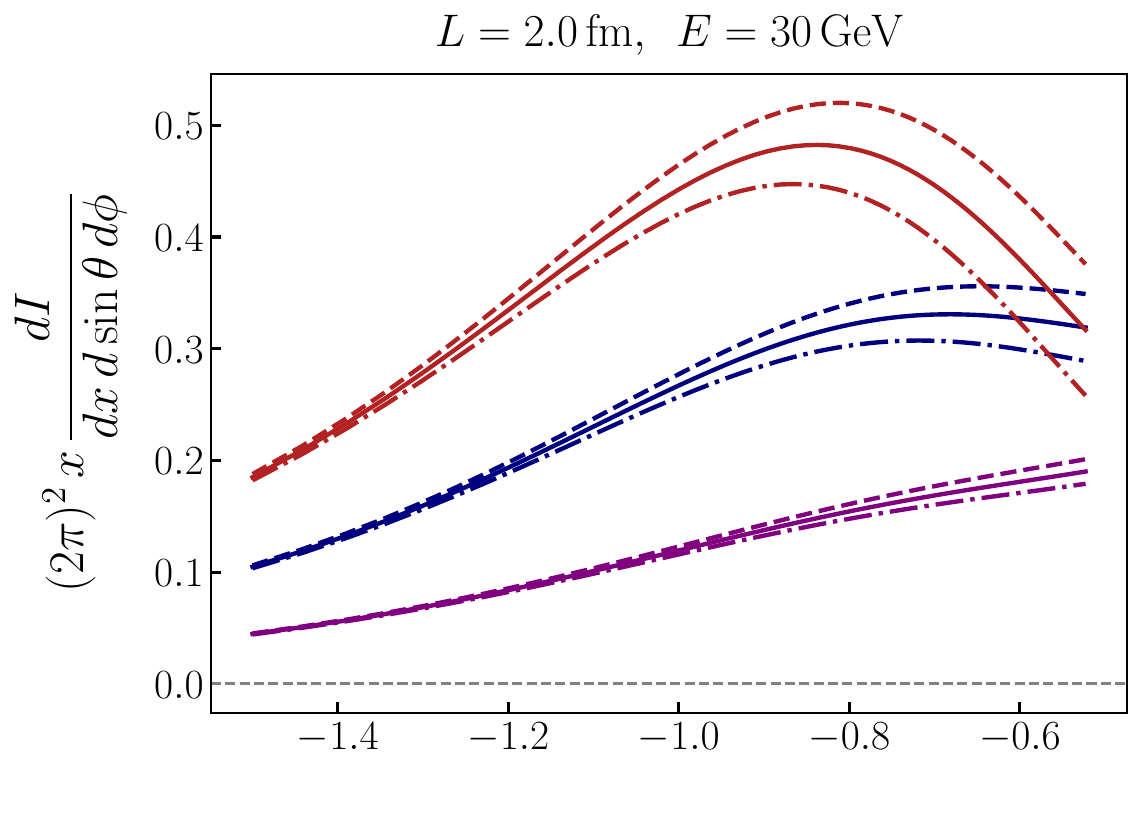}\hspace{0mm}
    \includegraphics[height=0.19\textheight]{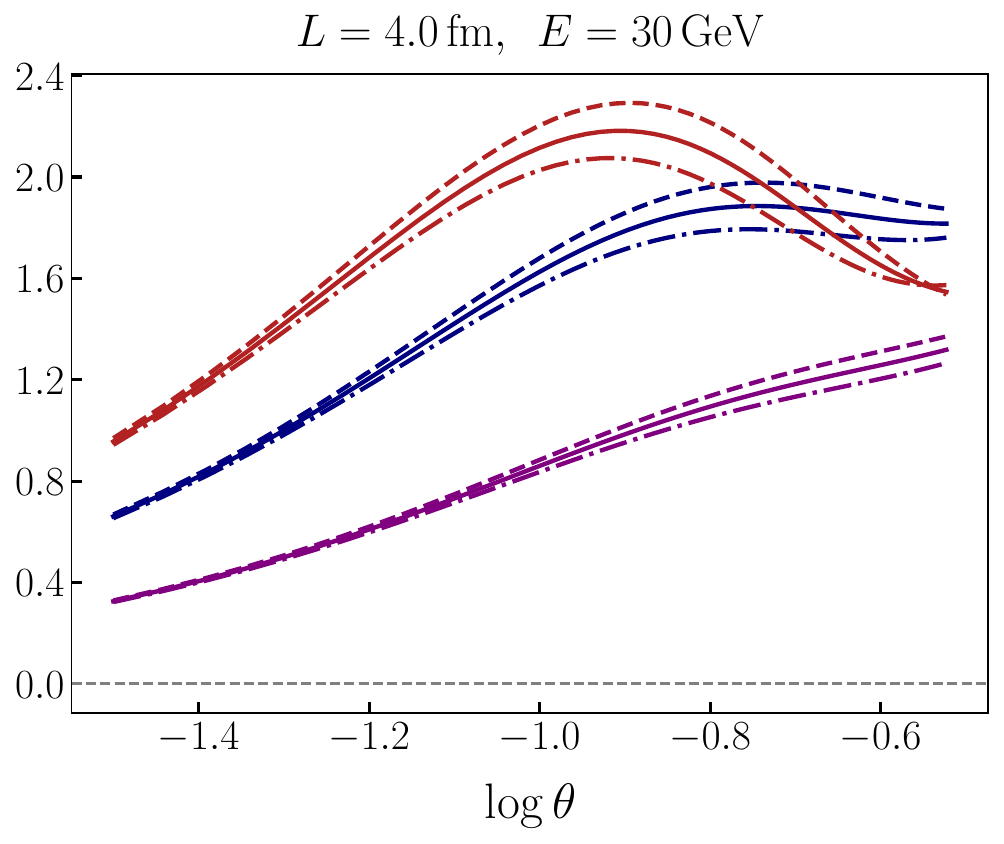}\hspace{0mm}
    \includegraphics[height=0.19\textheight]
    {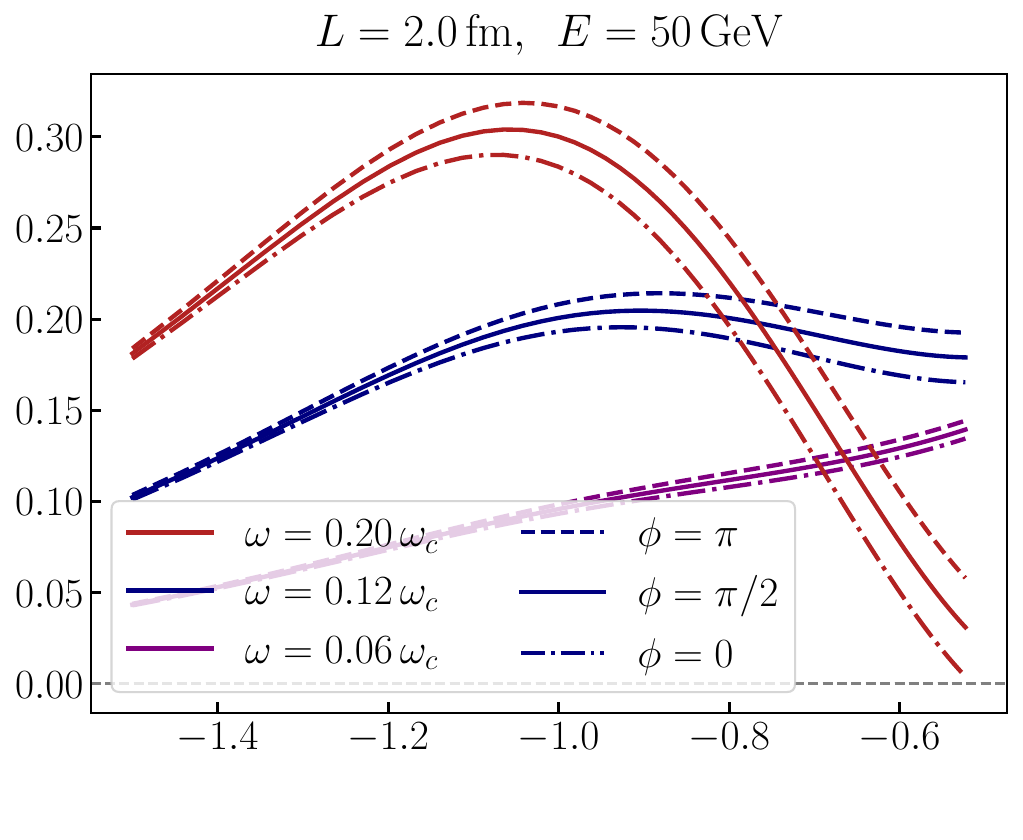}\
    \caption{The medium-induced spectrum off a heavy quark in flowing matter, given by Eq.~\eqref{eq:spec_final}, is shown for three gluon energies, $\omega = 0.06\,\omega_c$, $\omega = 0.12\,\omega_c$ and $\omega = 0.20\,\omega_c$, as a function of the gluon emission angle with respect to the heavy quark direction $\sin\theta\simeq\sqrt{2}\frac{|\k|}{\o}$. For each gluon energy, three different curves are plotted: the solid line shows the spectrum with $\k$ and $\u$ orthogonal ($\phi=\frac{\pi}{2}$) in the transverse plane, while the dash-dotted and dashed lines correspond to them being parallel ($\phi =0$) and anti-parallel ($\phi =\pi$) respectively. The magnitude of the transverse flow and the heavy quark mass are fixed at $|\u| =0.7$ and $m_{Q}=4$ GeV, as well as the jet quenching parameter $\hat{q}_0 = 1\,\text{GeV}^2\cdot\text{fm}^{-1}$, while $L$ and $E$ are varied in the three different panels.}
    \label{fig:spectrums}
\end{figure*}

The explicit form of the spectrum in Eq.~\eqref{eq:spec_final} neatly illustrates the emergence of a new effect, which we refer to as the \textit{directional dead-cone}. Examining the phase factors entering the expression, one readily notices a \textit{directional} contribution originating from the mass-dependent coupling of the heavy quark to the flow:
\begin{align}
    \Theta_{\rm dc}^2 (\k\cdot \u) = \theta_{\rm dc}^2  \left(1-\frac{\k\cdot \u}{u^-\omega}\right)\, .
\end{align}
The dependence of the resulting dead-cone angle on the kinematics of the emitted gluon is well expected, since no other vector structures are available to combine with $\u$.\footnote{Here we neglect terms that scale parametrically as $\theta_{\rm dc}^4$, since they lie beyond the accuracy of our present approximations.} Importantly, the dead-cone now acquires an anisotropic character: it is minimal (maximal) when the gluon's transverse momentum is (anti-)parallel with respect to the transverse flow field. This anisotropy produces an elliptic pattern in the medium-induced radiation inside the dead-cone, as illustrated in Fig.~\ref{fig:dead-cone_cartoon}. It is worth stressing that, although reminiscent of azimuthal modulations of the soft-gluon spectrum~\cite{Sadofyev:2021ohn,Barata:2023qds,Kuzmin:2023hko,Kuzmin:2024smy,Barata:2023zqg} and quark-antiquark antenna production \cite{Barata:2024bqp} in evolving and/or structured media, the underlying mechanism here is different: even if the medium-induced radiation were emitted isotropically, the suppression associated with virtuality-driven 
emissions would still generate a non-trivial dead-cone geometry.

In Fig.~\ref{fig:spectrums} we evaluate Eq.~\eqref{eq:spec_final} as a function of the emission angle of the soft gluon, defined as $\sin{\theta}\simeq\sqrt{2}\frac{|\k|}{\o}$. The three colors correspond to different gluon energies, $\omega = 0.06\,\omega_c$, $\omega = 0.12\,\omega_c$ and $\omega = 0.20\,\omega_c$, given in terms of the critical medium energy $\omega_c\equiv\frac{1}{2}\hat{q}_0 L^2$, which corresponds to the energy of gluons with formation time of order $L$. For each energy, we show three curves depending on the relative angle $\phi$ between $\k$ and $\u$: the solid line corresponds to the orthogonal configuration ($\phi=\frac{\pi}{2}$), while the dash–dotted and dashed lines correspond to the parallel ($\phi=0$) and anti-parallel ($\phi=\pi$) cases, respectively. The three panels further distinguish different choices of $L$ and $E$, while the magnitude of the flow, the mass of the heavy quark and the jet quenching parameter remain fixed at $|\u| =0.7$, $m_{Q}=4$ GeV and $\hat{q}_0 = 1\,\text{GeV}^2\cdot\text{fm}^{-1}$ respectively. These results show that gluon production (anti-)parallel to the flow is suppressed (enhanced), indicating the sensitivity of \textit{bremsstrahlung} to the medium evolution. The separation between the two curves grows with increasing $|\u|$ and $m_Q/E$, in accordance with Eq.~\eqref{eq:spec_final}. The relative effect appears to be stronger for shorter propagation lengths, although the spectrum depends on $\hat{q}_0$ and $L$ in a non-linear manner.

\vspace{0.25cm}
\noindent\emph{\textbf{Conclusion and discussion:}} We have presented the first calculation of the evolution of high-energy heavy quarks in a dense and flowing QCD medium. This reveals two new physical effects arising from the mass-dependent coupling of the heavy quark to the medium flow. The first is an emergent tensorial jet-quenching coefficient $\hat q_{ij}$, which leads to anisotropic momentum broadening. The second is the appearance of a \textit{directional dead-cone}, which, within the regime where medium-induced radiation populates the dead cone, produces an anisotropic energy distribution around the heavy quark. At the level of accuracy considered here, both effects are absent for light quarks, making heavy quarks particularly well suited as probes of the medium evolution. 

The current study focuses on purely theoretical/conceptual aspects of heavy quark evolution in flowing matter. A connection between our results and phenomenology would be natural in the context of anisotropic jet substructure and hard parton $v_n$ studies, {\it c.f.} with recent discussions on similar effects for light quark jets \cite{Antiporda:2021hpk,Barata:2023zqg,Bahder:2024jpa,Ke:2024emw,Sengupta:2025kdr}. Another promising possibility would be to utilize the jet-tree based observables~\cite{Cunqueiro:2022svx}, augmented with an azimuthal variable, to probe the directional effects. However, one could expect that such an approach might be sensitive to background contamination from the flowing bulk, which would obscure the physics associated to medium induced jet modifications. Such issues might be easily mitigated for, e.g., heavy hadron production, for which the background contamination is minimal, see~\cite{Barata:2025uxp} for a recent discussion. We leave such aspects for future work.

\noindent\emph{\textbf{Acknowledgments:}} We are grateful to J. M. Silva for multiple discussions on the numerical realization of the in-medium spectrum. This material is based upon work supported by the U.S. Department of Energy, Office of Science, Office of Nuclear Physics under grant Contract Number DE-SC0011090. This work is supported by the European Research Council under project ERC-2018-ADG-835105 YoctoLHC; by Maria de Maeztu excellence unit grant CEX2023-001318-M and project  PID2023-152762NB-I00 funded by MICIU/AEI/10.13039/501100011033; and by ERDF/EU. It has received funding from Xunta de Galicia (CIGUS Network of Research Centres). The work of AVS is supported by Fundação para a Ciência e a Tecnologia (FCT) under contract 2022.06565.CEECIND and by the Basque Government through grant IT1628-22. AVS would also like to acknowledge support from Ikerbasque, Basque Foundation for Science.

\bibliographystyle{JHEP-2modlong.bst}

\bibliography{references.bib}

\clearpage
\onecolumngrid  
\appendix

\section*{Supplemental material}

\renewcommand{\theequation}{A\arabic{equation}}
\setcounter{equation}{0}

\subsection{Transverse momentum broadening}

In this section we detail the derivation of the momentum broadening of a heavy quark interacting with flowing QCD matter, closely following \cite{Barata:2022krd,Andres:2022ndd,Barata:2023qds}. For simplicity, we ignore the energy-suppressed spin effects, treating the heavy quark as a scalar particle in the fundamental representation, see e.g. \cite{Sadofyev:2021ohn} for a discussion. We use the light-cone coordinates $x^\mu=(x^+,x^-,\x)^\m$  for convenience, with the $\pm$ components defined as $x^\pm = \frac{x^0 \pm x^3}{\sqrt{2}}$, and $p\cdot x = p^- x^+ + p^+ x^-  - \p\cdot\x$. 

Following \cite{Sadofyev:2021ohn}, we model the QCD matter by a classical stochastic color field generated by moving quasi-particle sources, given in Eq.~\eqref{eq: field def}.
A heavy quark propagating inside QCD matter will experience an arbitrary number of tree-level gluon exchanges with the bulk, as depicted in Fig.~\ref{fig:diagram_broad}. With the background field formalism at hand, one can write the amplitude with $N$ gluon exchanges as
\begin{align}\label{eq:BroadeningAmplitude}
     i\cM_N(p)  &=  \prod_{n=1}^{N}\Big[\int_{p_n}it^{a_n}g A^{a_n}(p_{n+1}-p_n)\cdot(p_{n+1}+p_n)\frac{i}{p_n^{2}-m^2_Q+i\epsilon}\Big]J(p_1)
\end{align}
with $t^a$ being the heavy quark color generator and $p = p_{N+1}$ its final momentum.
This amplitude can be significantly simplified in the high-energy limit, where the light-cone energy of the final-state quark is taken to be larger than the characteristic transverse and medium scales, $E\gg \bot,\mu$, and the ordering $|\p|/E\ll m_{Q}^2/E^2 <1$ is assumed to neglect certain kinematic corrections simplifying the calculation further. As extensively discussed in the literature, see \textit{e.g.} \cite{Sadofyev:2021ohn,Barata:2022krd,Andres:2022ndd,Barata:2023qds}, we assume that $\mu \Delta x^+ \gg 1$ with $\Delta x^+$ the characteristic light-cone distance between color sources (determined by $\rho^a$). Thus, the poles coming from the screened in-medium potentials are exponentially suppressed, and $p_n^{-}$ integrations are governed solely by the poles of the propagators. Thus, by summing over the number of gluon exchanges with the medium, one obtains the resummed amplitude of the process, which reads
\begin{align}
     i\cM(p)&=\sum_{N=0}^\infty iM_N(p)=
     \sum_{N=0}^\infty \prod_{n=1}^{N}\Bigg[i\left(1+\frac{u^+}{u^-}\frac{m_Q^2}{E^2}\right)\int_{\x_n,x^+_n}\int_{\p_n}e^{-i(\p_{n+1}-\p_n)\cdot \x_{n}}e^{-i\frac{m^2_{Q}}{E^2}\frac{(\p_{n+1}-\p_n)\cdot\u}{2 u_-}x^+_{n}} \notag
     \\
     \notag &\hspace{4.5cm}\times t^{a_n}v\left((\p_{n+1}-\p_n)^2+\frac{m_Q^2}{E^2}\Big[\frac{\u\cdot (\p_{n+1}-\p_n)}{u^-}\Big]^2\right) \rho^{a_n}(\x_n, x_n^+)\theta_{n,n+1}\Bigg]J(\p_1)     
     \\
     \notag &
    =\int_{\x}e^{-i\p\cdot \x} \sum_{N=0}^\infty \prod_{n=1}^{N}\Bigg[i\int_{x^+_n}t^{a_n} v^{a_n}\left(\x-\frac{m_Q^2}{E^2}\frac{\u}{2 u_-}x^+_{n},x^+_{n}\right)\theta_{n,n+1}\Bigg]J(\x) 
    \\
    &= \int_{\x} e^{-i\p\cdot \x} \, W^{[u]}(\x;L,0) \, J(\x)  \,,
    \label{eq:BroadeningAmplitudeAppendix}
\end{align}
where we use the shorthand notations $\int_x \equiv \int dx$ (and $\int_p \equiv \int \frac{dp}{2\pi}$) and  $\theta_{n,n+1} = \theta(x^+_{n+1}-x^+_{n})$, with the modified scattering potential $v^a$ introduced in Eq.~\eqref{eq:vamodified}.
The final-state parton distribution follows straightforwardly from the squared and averaged resummed amplitude, as in the massless case, yielding
\begin{align}\label{eq:BroadCrossSection}
    E\frac{d\mathcal{N}}{d^2\p \,dE} =\frac{1}{2(2\pi)^3} \int_{\y,\bar \y}\left\langle\,\cW^{[u]}(\y;L,0)\cW^{[u]\dagger}(\bar \y;L,0)\right\rangle e^{-i\p\cdot(\y-\bar \y)}J(\y)J^{\dagger}(\bar \y)\,.
\end{align}
The final-state parton distribution is governed entirely by a two-point function, as in the massless case, with flow effects encoded in the modifications of the Wilson lines. To evaluate this correlator, one needs to specify the average over the stochastic field configurations. Assuming Gaussian white-noise statistics for simplicity, one finds that only the pairwise correlators of the field survive, and these can be written in terms of the in-medium color sources as
\begin{equation}\label{eq: <rho rho>}
    \left<\rho^a(\x,x^+) \rho^b(\y,y^+)\right> = \frac{1}{2C_{A}} \d^{ab} \d^{(2)}(\x-\y) \d(x^+-y^+) \, \rho(\x,x^+) \,.
\end{equation}
where the in-medium sources are taken to be in the fundamental representation, and $\rho(\x,x^+)$ denotes their number density. The locality of Eq.~\eqref{eq: <rho rho>} in the $+$ component allows one to build and solve the correlator evolution equation, closely following the massless case \cite{Andres:2022ndd}. The resulting solution is given in Eq.~\eqref{eq:amp_B_WW}, with $\rho$ treated as constant. 

Transverse momentum broadening is often characterized by $\hat{q}$, which measures the squared transverse momentum gained by the jet per unit length in the matter. This parameter can be extracted directly from Eq.~\eqref{eq:BroadCrossSection}, leading to
\begin{align}\label{eq: qhat def}
    \hat{q}_{ij} = \frac{\partial}{\partial L}\left\langle \p_i \p_j\right\rangle \equiv \frac{\partial}{\partial L}\frac{\int_{\p}E\frac{d\mathcal{N}}{d^2\p \,dE} \,\p_i\p_j}{\int_{\p}E\frac{d\mathcal{N}^{(0)}}{d^2\p \,dE}} = - \frac{\partial}{\partial L}\Big[\na_{(y-\bar{y})_i}\na_{(y-\bar{y})_j} \left\langle W^{[u]}(\y;L,0)W^{[u]\dag}(\bar \y;L,0)\right\rangle\Big]_{\bar{\y}=\y}\,,
\end{align}
where $E\frac{d\mathcal{N}^{(0)}}{d^2\p \,dE}= \frac{1}{2(2\pi)^3}|J(E,\p)|^2$ is the initial distribution. Using the model potential  $v(\q^2)\equiv-\frac{g^2}{\q^2+\mu^2}$ together with Eq.~\eqref{eq:amp_B_WW}, one readily arrives at Eq.~\eqref{eq:qhat_flowing}. In the massless limit, $\hat{q}_{ij}$ is diagonal, reducing to $\hat{q}_0\equiv \frac{C_F}{2N_c} \rho \int_\q \q^2 v^2(\q^2) $. 

\subsection{Medium-induced soft gluon radiation}

\noindent\emph{\textbf{Resummation at the amplitude level:}} In this section, we briefly review the derivation of the resummed amplitude for gluon emission off a heavy quark traversing a flowing medium. Using the background-field model introduced in the previous section and treating the heavy quark as a scalar in the fundamental representation, the amplitude with $N_p$ insertions in the initial heavy-quark  line, $N_l$ insertions in the final heavy-quark line, and $N_k$ insertions in the gluon line, shown in Fig.~\ref{fig:diagram_M_final}, can be written as
\begin{align}\label{eq:AmplitudeStart}
    i\mathcal{R}_{N_p N_l N_k} &=\prod_{n=1}^{N_p} \left[(-1)\int_{p_n,p^+_n} t^{a_n} g A^{a_n}(p_{n+1}-p_n)\cdot (p_{n+1}+p_{n})
    \frac{1}{p_n^2-m_Q^2+i\epsilon}\right] \notag 
    \\ & \times \int dx^+_s \frac{d^4p_s}{(2\pi)^4}\,  (-1)\,gt^{b_1}\frac{(p_s+l_{1})_{\mu_1}}{p_s^2-m_Q^2+i\epsilon}\,(2\pi)^3 \delta^{(3)}(p_s-l_{1}-k_{1}) e^{ix^+_s(l^-_{1}+k^-_{1}-p_s^-)} J(p_1)\notag
    \\ & \times \prod_{r=1}^{N_k} \left[\left(-\frac{i}{g}\right)\int_{k_r} \frac{N^{\mu_r \nu_r}_{k_r}}{k_r^2+i\epsilon} 
    \Gamma^{b_r b_{r+1} c_r}_{\nu_r \mu_{r+1} \lambda_r}(k_r,-k_{r+1})\,gA^{c_r \lambda_r}(k_{r+1}-k_r)\right]\epsilon^{*\mu_{N_k+1}}_k\notag
    \\ & \times  \prod_{m=1}^{N_l} \left[(-1)\int_{l_m} t^{d_m}gA^{d_m}(l_{m+1}-l_m)\cdot(l_{n+1}+l_{n})
    \frac{1}{l_m^2-m_Q^2+i\epsilon}\right],
\end{align}
where $\Gamma ^{abc}_{\a\beta\gamma}$ is the three-gluon vertex, $p_{N_p+1}\equiv p_s$, $l_{N_l+1}\equiv l$, and $k_{N_k+1}\equiv k$. We have also partially rewritten the momentum-conservation delta function in the emission vertex as an integral for later convenience. The integrations in \eqref{eq:AmplitudeStart} should be understood as acting on the entire expression. This amplitude can be considerably simplified in the large-energy limit, where $E\gg \bot,\mu$. For simplicity, we work in the soft-gluon limit, assuming $\o=x E \ll E$. Since evaluating the gluon-emission amplitude is more involved, we only keep those subeikonal terms that are enhanced by powers of $x^{-1}$ or $m_Q$, as well as terms scaling with the medium length up to $\mathcal{O}\left(\left(\frac{\bot^2}{x E } \, x^+\right)^n\right)$ and $\mathcal{O}\left(\left(\frac{m_Q^2}{E^2}  \, \bot \, x^+\right)^n\right)$, \textit{i.e.} the LPM phase and their leading mass correction, while neglecting all contributions beyond that.

We further perform the $l^+_m$-integration using the delta functions from the field definition Eq.~\eqref{eq: field def}. As in the previous section, we consider an extended and sufficiently dilute medium such that $\mu \Delta x^+ \gg 1$, so that in the $l_m^-$ integrations we collect only the poles originating from the propagators. Thus,
\begin{align} \label{eq: integration final quark leg}
     \notag &  \prod_{m=1}^{N_l} \left[(-1)\int_{l_m} t^{d_m}gA^{d_m}(l_{m+1}-l_m)\cdot(l_{n+1}+l_{n})
    \frac{1}{l_m^2-m^2_{Q}+i\epsilon}\right] \, \notag
    \\ & = \prod_{m=1}^{N_l} \Big[ i \int_{\l_m, \x_{m},\,x_m^+} t^{d_m} \rho^{d_m}(\x_m, x_m^+) e^{-i(\l_{m+1}-\l_m)\cdot \x_{m}}  v(\l_{m+1}-\l_m) \, \notag
    \\ & \hspace{3 cm}\times e^{- i \frac{m^2_Q}{E^2} \frac{(\l_{m+1}-\l_m)\cdot \u}{2 u^-}\, x_m^+} \, \theta_{m,m-1} \Big] \, ,
\end{align}
where we account for the $l_1^-$-dependent vertex phase by identifying $x^+_0 \equiv x^+_s$. Summing over $N_l$, we obtain
\begin{align}
    & (p_s + l )_{\mu_1} + \sum_{N_l=1}^{\infty} \prod_{m=1}^{N_l} \Big[ i \int_{\l_m, \x_{m},x_m^+} t^{d_m} \rho^{d_m}(\x_m, x_m^+) e^{-i(\l_{m+1}-\l_m)\cdot \x_{m}}  v\left((\l_{m+1}-\l_m)^2\right) \, \notag
    \\ & \hspace{3 cm}\times e^{- i \frac{m_Q^2}{E^2} \frac{(\l_{m+1}-\l_m)\cdot \u}{2 u^-}\, x_m^+} \, \theta_{m,m-1} \Big] (p_s + l_1 )_{\mu_1} \, \notag
    \\ & \hspace{1cm} =  \int_{\l_1,\y}e^{- i \y \cdot (\l-\l_1)} \, W^{[u]}(\y;\infty,x^+_s) (p_s + l_1 )_{\mu_1}
\end{align}
where, within the accuracy, the potential Eq.~\eqref{eq:vamodified} reduces to the massless case and the only flow dependence of the Wilson line comes from its argument shift. 

Analogously, the initial quark line can be simplified to
\begin{align}
    & J(p_s) + \sum_{N_p=1}^{\infty}\prod_{n=1}^{N_p} \left[(-1)\int_{p_n} t^{a_n} g A^{a_n}(p_{n+1}-p_n)\cdot (p_{n+1}+p_{n})
    \frac{1}{p_n^2-m_Q^2+i\epsilon}\right] J(p_1) \, \notag
    \\
    &=\int_{\p_1,\x} e^{-i \x \cdot (\p_s-\p_1)}W^{[u]}(\x;x^+_s, 0) \, J(p_1) \, ,  
\end{align}
where we have also integrated over $p_s^-$, picking up the propagator pole and setting $x^+_{N_p}<x_s^+$, and introduced $x^+_0=0$.

Finally, we turn to the gluon leg. The gluon is taken to be energetic and collinear, satisfying $\o=xE \gg |\k|$ with $x \ll 1$. For simplicity, we work in the axial gauge $u \cdot A \propto u^2 = 0 $, which makes both the gluon propagator and the polarization vector transverse to the flow vector. With this choice, the former reduces to Eq.~\eqref{eq: gluon prop} and the latter to Eq.~\eqref{eq: pol vec}. After some algebra, the gluon line simplifies considerably, giving
\begin{align} \label{eq: gluon leg denominators simplified}
    & \prod_{r=1}^{N_k} \left[\left(-\frac{i}{g}\right)\int_{k_r} \frac{N^{\mu_r \nu_r}(k_r)}{k_r^2+i\epsilon} 
    \Gamma^{b_r b_{r+1} c_r}_{\nu_r \mu_{r+1} \sigma_r}(k_r,-k_{r+1})\,gA^{c_r \sigma_r}(k_{r+1}-k_r)\right]\epsilon^{*\mu_{N_k+1}}_k \, \notag
    \\ &= \prod_{r=1}^{N_k} \left[-\int_{\k_r, k_r^-,x_r} (T^{c_{r}} )_{b_{r+1} b_{r}} \, \rho^{c_r}(\x_r,x^+_r)  v\left(-(k_{r+1}-k_r)^2\right) \frac{2 k_r \cdot u}{k_r^2 + i \e} \frac{1}{u^-} \, e^{-i \x_r(\k_{r+1}-\k_r)}e^{i x^{+}_r(k^-_{r+1}-k^-_r)}\right] \, \notag 
    \\ & \hspace{4cm} \times \Big[\epsilon^{*\mu}(k)-\frac{k_{1}\cdot \epsilon^*(k)}{k\cdot u} u^{\mu}\Big] \, ,
\end{align}
where $T^{a}$ is the generator of the adjoint representation, and  we have integrated over each $k_r^+$ using the delta function inside the field. Similarly to the quark legs, we can integrate \eqref{eq: gluon leg denominators simplified} over the $k_r^-$ by residues, obtaining
\begin{align}\label{eq: Integrated gluon line}
    \prod_{r=1}^{N_k} & \left[i\int_{\k_r,x_r} (T^{c_{r}} )_{b_{r+1} b_{r}} \, \rho^{c_r}(\x_r,x^+_r)  v\left((\k_{r+1}-\k_r)^2\right)  \, e^{-i \x_r(\k_{r+1}-\k_r)}e^{i x^{+}_r \frac{\k_{r+1}^2-\k_{r}^2}{2 \o} } \theta_{r,r-1}\right] \, \notag 
    \\ & \hspace{4cm} \times \Big[\epsilon^{*\mu}(k)-\frac{\tilde{k}_{1}\cdot \epsilon^*(k)}{k\cdot u} u^{\mu}\Big] \, ,
\end{align}
where the tilde indicates that the minus components of the momenta are fixed by the corresponding poles, and we again set $x^+_0 \equiv x^+_s$. The first line of \eqref{eq: Integrated gluon line} receives no corrections from flow or mass effects, and therefore has the same structure as in the static, massless case. It can be straightforwardly resummed to 
\begin{align}\label{eq: Fully resummed gluon line}
    \delta^{b_{N_k+1} b_{1}} \e^{*\mu_1}(k) &+ \sum_{N_k=1}^{\infty} \prod_{r=1}^{N_k} \left[i\int_{\k_r,x_r} (T^{c_{r}} )_{b_{r+1} b_{r}} \, \rho^{c_r}(\x_r,x^+_r)  v\left((\k_{r+1}-\k_r)^2\right)  \, e^{-i \x_r(\k_{r+1}-\k_r)}e^{i x^{+}_r \frac{\k_{r+1}^2-\k_{r}^2}{2 \o} } \theta_{r,r-1}\right] \, \notag 
    \\ & \hspace{3cm} \times \Big[\epsilon^{*\mu}(k)-\frac{\tilde{k}_{1}\cdot \epsilon^*(k)}{k\cdot u} u^{\mu}\Big] \, \notag
    \\ & = \lim_{x^+_f \rightarrow \infty} \int_{\k_1} e^{-ix_s^+ \frac{\k_1^2}{2\o}} \, e^{ix_f^+ \frac{\k^2}{2\o}} \, \cG^{b_{N_k+1} b_{1}}(\k,x^+_f;\k_1,x^+_s) \Big[\epsilon^{*\mu}(k)-\frac{\tilde{k}_{1}\cdot \epsilon^*(k)}{k\cdot u} u^{\mu}\Big] \, ,
\end{align}
where we have introduced the single-particle propagator, which can be expressed as 
\begin{align}
    \cG(\x_f,x^+_f;\x_i,x^+_i) = \int^{\x_f}_{\x_i} \cD\r \exp\Big\{i\frac{\o}{2} \int^{x^+_f}_{x^+_i} d\t \dot{\r}^2 \Big\} \, \cP\exp\Big\{i\int^{x^+_f}_{x^+_i} d\t  \, T^c \, v^c(\r(\t),\t) \Big\} \, ,
\end{align}
and the auxiliary variable $x^+_f$ is introduced to simplify the definition of $\cG$ in \eqref{eq: Fully resummed gluon line}.

After integrating over the minus components of the intermediate momenta as detailed above, the Fourier phase arising from momentum conservation in the emission vertex becomes
\begin{align}
    x^+_s \left(\tilde{l}^-_1+\tilde{k}^-_1 - \tilde{p}^-_s\right) = x^+_s \left(\frac{m^2_{Q}x^2 + \k_1^2}{2\o}-\frac{m^2_{Q}}{E^2}\frac{(\k-\k_1)\cdot\u}{2u^-}\right) \, ,
\end{align}
from which we can readily identify the dead-cone angle contribution $ m^2_{Q}x^2 =\theta_{dc}^2 \o^2$, the only non-vanishing term in the static-matter limit, together with its novel correction arising from the interplay between flow and mass. Finally, the resummed amplitude can be written as 
\begin{align}\label{eq: resummed amplitude not simplified}
    i \cR \simeq & i \frac{g}{\o} \lim_{x^+_f \rightarrow \infty} \int_{x_s^+,\k_1,\x}  J(\y) \,  e^{-i \y\cdot(\l+\k_1)} \, e^{i x^+_s \left(\frac{m^2_{Q}x^2}{2\o}-\frac{m^2_{Q}}{E^2}\frac{(\k-\k_1)\cdot\u}{2u^-}\right)}  \, \notag
    \\ & \times  \beps \cdot
    \k_1 \, W^{[u]}(\y;\infty,x^+_s) \, t^{b_1} \, W^{[u]}(\y;x^+_s,0) \, e^{ix_f^+ \frac{\k^2}{2\o}} \, \cG^{b_f b_{1}}(\k,x^+_f;\k_1,x^+_s)  \, ,
\end{align}
where the color indices of the Wilson lines and of the fundamental $SU(N_c)$ generator are omitted for simplicity.

\vspace{0.5cm}
\noindent\emph{\textbf{Final state parton distribution:}} 
Squaring \eqref{eq: resummed amplitude not simplified} and integrating over the final quark transverse momentum $\l$, we fix the positions of the Wilson lines to be the same in the amplitude and its conjugate.  Following the static, massless case, we further express the full distribution as the real part of the contribution with $\bar{x}^+_s > x^+_s$. The fundamental heavy-quark Wilson lines in the presence of flowing matter can still be combined into a single adjoint Wilson line, and we finally find 
\begin{align} \label{eq: initial squared amplitude}
    & 2(2\pi)^3 \o E \frac{d\cN}{d\o dE d^2\k} =  \lim_{x^+_f\rightarrow\infty} \frac{\alpha_s}{N_c \, \o^2} \re \int_0^\infty d\bar{x}^+_s \int_0^{\bar{x}^+_s} dx^+_s \int_{\y,\k_1,\bar{\k}_1,\l} \, |J(\y)|^2 \, e^{-i\y\cdot(\k_1-\bar{\k}_1)} \, \notag
    \\ & \times e^{i x^+_s \left(\frac{m^2_{Q}x^2}{2\o}-\frac{m^2_{Q}}{E^2}\frac{(\k-\k_1)\cdot\u}{2u^-}\right)} \, e^{-i \bar{x}^+_s \left(\frac{m^2_{Q}x^2}{2\o}-\frac{m^2_{Q}}{E^2}\frac{(\k-\bar{\k}_1)\cdot\u}{2u^-}\right)} \,  \beps \cdot
    \k_1 \,  \bar{\beps} \cdot
    \bar{\k}_1  \, \notag
    \\ & \times \left< \cG^{b c}(\k,x^+_f;\l,\bar{x}^+_s) \cG^{\dag a b}(\k,x^+_f;\bar{\k}_1,\bar{x}^+_s)    \,  \cG^{c d}(\l,\bar{x}^+_s;\k_1,x^+_s) W_A^{[u]\dag d a}(\y;\bar{x}^+_s,x^+_s) \right>  \, ,
\end{align}
where  $W_A^{[u]}$ is the adjoint Wilson line, and we have relabeled the adjoint indices for simplicity. 

Because the medium averages are local in $x^+$, the average of a product of propagators with no common support, such as the one in Eq.~\eqref{eq: initial squared amplitude}, reduces to a product of independent averages. Furthermore, the color triviality of Eq.~\eqref{eq: <rho rho>} allows one to simplify the color structure of each average in Eq.~\eqref{eq: initial squared amplitude}, writing the full process in terms of the following two colorless objects,
\begin{align}
    S_2(\k,\k,x^+_f;\l,\bar{\k}_1,\bar{x}^+_s) & = \frac{1}{N_c^2-1} \left<\cG^{b a}(\k,x^+_f;\l,\bar{x}^+_s) \cG^{\dag a b}(\k,x^+_f;\bar{\k}_1,\bar{x}^+_s)\right> \, ,
    \\ \cK^{[u]}(\l,\y,\bar{x}^+_s; \k_1,\y,x^+_s) & = \frac{1}{N_c^2-1} \left<\cG^{cd}(\l,\bar{x}^+_s;\k_1,x^+_s) W_A^{[u]\dag d c}(\y;\bar{x}^+_s,x^+_s)\right> \, , 
\end{align}
known as the broadening kernel, $S_2$, which describes the evolution of the radiated gluon, and the emission kernel, $\cK$, which governs its formation. With these definitions, the final-state parton distribution takes the form
\begin{align} 
\label{eq:spectrum}
    & 2(2\pi)^3 \o E \frac{d\cN}{d\o dE d^2\k} =  \lim_{x^+_f\rightarrow\infty} \frac{\alpha_s}{N_c \, \o^2} \re \int_0^\infty d\bar{x}^+_s \int_0^{\bar{x}^+_s} dx^+_s \int_{\y,\k_1,\bar{\k}_1,\l} \, |J(\y)|^2 \, e^{-i\y\cdot(\k_1-\bar{\k}_1)} \, \notag
    \\ & \times e^{i x^+_s \left(\frac{m^2_{Q}x^2}{2\o}-\frac{m^2_{Q}}{E^2}\frac{(\k-\k_1)\cdot\u}{2u^-}\right)} \, e^{-i \bar{x}^+_s \left(\frac{m^2_{Q}x^2}{2\o}-\frac{m^2_{Q}}{E^2}\frac{(\k-\bar{\k}_1)\cdot\u}{2u^-}\right)} \,  \beps \cdot
    \k_1 \, \bar{\beps} \cdot
    \bar{\k}_1 \, \notag
    \\ & \times S_2(\k,\k,x^+_f;\l,\bar{\k}_1,\bar{x}^+_s) \, \cK^{[u]}(\l,\y,\bar{x}^+_s; \k_1,\y,x^+_s)  \, ,
\end{align}
where all modifications due to the medium flow are contained solely in the vertex phases and the emission kernel $\cK^{[u]}$.

\vspace{0.5cm}
\noindent\emph{\textbf{Medium induced gluon spectrum:}} Here, we focus on obtaining a closed form for the medium-induced gluon distribution, which requires further simplifying approximations for analytic treatment. We neglect the initial state effects by taking the broad source approximation, setting $|J(\y)|^2=f(E) \, \d^{(2)}(\y)$ and introducing the medium-induced gluon spectrum. The broadening kernel in Eq.~\eqref{eq:spectrum} is unmodified by the flow effects, and we use its well-known form, 
\begin{align}
    S_2\left(\k , \k, x^+_f; \x , \bar{\x}, \bar{x}^+_s\right) =e^{-i\k \cdot(\x-\bar \x)}  e^{-\int_{\bar{x}^+_s}^{x^+_s} d\t \, \cV(\x-\bar{\x},\t)} \, ,
\end{align}
where we have introduced dipole potential
\begin{align}
    \cV(\q,x^+) = - \frac{C_F}{2 N_c} \rho(x^+) \left(v^2(\q^2) - (2\pi)^2\delta^{(2)}(\q) \int_{\l} v^2(\l^2) \right)\, .
\end{align}

In turn, the evolution equation satisfied by the modified emission kernel reads
\begin{align} \label{eq: diff eq K}
    &\frac{\partial}{\partial \bar{x}^+} \cK^{[u]}(\x_f,\bar{x}^+; \x_i,x^+_s) \, \notag
    \\ & \hspace{2cm} = \left[ i \frac{\boldsymbol{\na}^2_{\x_f}}{2\o} - \cV\left(\x_f + \frac{m^2_{Q}}{E^2} \frac{\u}{2u^-} \bar{x}^+; \bar{x}^+\right)\right]  \cK^{[u]}(\x_f,\bar{x}^+; \x_i,x^+_s) \, .
\end{align}
The solution to this Schrödinger-like differential equation can be formally written as a path integral, resembling the static case but with the shifted in-medium potential. It can be evaluated analytically only for specific $\cV$. A common choice is the so-called harmonic approximation, which renders the path integral Gaussian, and we rely on it here. By transforming coordinates, we partially shift the arguments into the phases, relating the kernel in flowing matter to its static counterpart. Doing so, we find
\begin{align}
    &\cK^{[u]}\left(\x_f,\bar{x}^+_s; \x_i,x^+_s\right) \equiv \cK^{[u]}\left(\x_f,\boldsymbol{0},\bar{x}^+_s; \x_i,\boldsymbol{0},x^+_s\right) = e^{i \frac{\o}{2} \left(\frac{m^2_{Q}}{E^2} \frac{\u}{2u^-}\right)^2 (\bar{x}^+_s-x^+_s)} \, \, \notag
    \\ & \hspace{1cm}  \times  e^{-i \frac{m^2_{Q}}{E^2}\o\frac{\u}{2u^-} \cdot \left(\x_f - \x_i + \frac{m^2_{Q}}{E^2} \frac{\u}{2u^-}(\bar{x}^+_s-x^+_s)\right)} \cK\left(\x_f + \frac{m^2_{Q}}{E^2} \frac{\u}{2u^-}\bar{x}^+_s ,\bar{x}^+_s; \x_i + \frac{m^2_{Q}}{E^2} \frac{\u}{2u^-} x^+_s ,x^+_s\right) \, ,
\end{align}
where $\cK$ is the emission kernel in the static limit:
\begin{align}\label{eq:KLO}
    	& \cK(\x,x^+_2;\y,x^+_1) =\frac{\Omega\,\omega}{2\pi i \sin\left(\Omega (x^+_2-x^+_1)\right)}\exp \left\{ \frac{i\Omega\,\omega}{2\sin\left(\Omega (x^+_2-x^+_1)\right)}\Big[(\x^2+\y^2)\cos \left(\Omega (x^+_2-x^+_1)\right)\,-2\x\cdot\y\Big]\right \}\,,
\end{align}
with $\Omega=\frac{1-i}{2}\sqrt{\frac{\hat q_0}{\omega}}$. We further sum over the final gluon polarizations and replace the initial momenta of the gluon by transverse gradients. Integrating by parts, we find that the medium-induced gluon spectrum reads
\begin{align}
    (2\pi)^2 \o \frac{dI}{d\o d^2\k} = &  \lim_{x^+_f\rightarrow\infty} \frac{\alpha_s}{N_c \, \o^2} \re \int_0^\infty d\bar{x}^+_s \int_0^{\bar{x}^+_s} dx^+_s \int_{\y} \, e^{-i \frac{m^2_{Q}x^2}{2\o}  \left(1-\frac{\k\cdot\u}{u^- \o} - \frac{m^2_{Q}}{E^2} \frac{\u^2}{4 u^{-2}}\right) \left(\bar{x}^+_s - x^+_s \right)} \, \notag
    \\ & \hspace{0cm} \times  e^{-i \frac{m^2_{Q}}{E^2} \o \frac{\u \cdot (\y-\x)}{2u^-}}   \, S_2(\k,\k,x^+_f;\y,\r,\bar{x}^+_s) \, \left(\boldsymbol{\na}_{\y} \cdot  \boldsymbol{\na}_{\x} \right) \,\cK(\y,\bar{x}^+_s; \x,x^+_s) \Big|_{\r=\x = \boldsymbol{0}}   \, .
\end{align}

The light-cone extension of the matter $L$ is usually assumed to be large but finite. It is therefore convenient to split the full spectrum into three  contributions corresponding to the integration regions: $\bar{x}^+_s < L$, $x^+_s <L<\bar{x}^+_s $, and $L<x^+_s$, commonly referred to as In-In, In-Out and Out-Out, respectively. For any point outside the matter, $\cV(\q,x^+>L)=0$, or equivalently $\hat{q}_0=0$, 
see \textit{e.g.} \cite{Barata:2023qds} for a detailed discussion. In the In-In region, one can evaluate the vertex derivatives and perform the integration over $\y$ explicitly, obtaining
\begin{align} \label{eq: dI In-In}
    &(2\pi)^2 \o \frac{dI^{\text{In-In}}}{d\o d^2\k} = \frac{2 \a_s C_F}{\o^2} \re \int_0^L d\bar{x}^+_s  \int_0^{\bar{x}^+_s} dx^+_s e^{-i \frac{m^2_{Q}x^2}{2\o}  \left(1-\frac{\k\cdot\u}{u^- \o} - \frac{m^2_{Q}}{E^2} \frac{\u^2}{4 u^{-2}}\right) \left(\bar{x}^+_s - x^+_s \right)} e^{-\frac{\left(\k + \frac{m^2_{Q}}{E^2} \o \frac{\u}{2u^-}\right)^2}{\hat{q}_0 \left(L  - \bar{x}^+_s\right) - 2 i \o \O \cot\left(\O \left(\bar{x}^+_s - x^+_s \right)\right) }}  \, \notag
    \\ & \hspace{0cm} \times \frac{8 \o^2 \O^2}{\sin\left(\O\left(\bar{x}^+_s - x^+_s \right)\right)^2} \frac{i \tfrac{1}{2}\hat{q}_0^2 \left(L -\bar{x}^+_s \right)^2 + \o \O \cot\left(\O \left(\bar{x}^+_s - x^+_s \right)\right) \left( \hat{q}_0 \left(L  - \bar{x}^+_s\right) + \left(\k + \frac{m^2_{Q}}{E^2} \o \frac{\u}{2u^-}\right)^2 \right)}{\left[i\hat{q}_0 \left(L  - \bar{x}^+_s\right) + 2 \o \O \cot\left(\O \left(\bar{x}^+_s - x^+_s \right)\right) \right]^3}\, .
\end{align}
Turning to the In-Out region, one can explicitly evaluate the vertex derivatives and integrate over the auxiliary coordinate $\y$. Due to the simplification of the spectrum  outside the matter, the second $x^+$-integral can be taken analytically, resulting in 
\begin{align} \label{eq: dI In-Out}
    &(2\pi)^2 \o \frac{dI^{\text{In-Out}}}{d\o d^2\k} = \frac{4 \a_s C_F}{\o} \re \int_0^L d\bar{x}^+_s e^{-i \frac{m^2_{Q}x^2}{2\o}  \left(1-\frac{\k\cdot\u}{u^- \o} - \frac{m^2_{Q}}{E^2} \frac{\u^2}{4 u^{-2}}\right) \left(L - x^+_s \right)} \, \notag 
    \\ & \hspace{3cm} \times \frac{-i}{\k^2 + m^2_{Q}\x^2} \, \frac{\left(\k + \frac{m^2_{Q}}{E^2} \o \frac{\u}{2u^-}\right)^2}{\cos\left(\O\left(L-x^+_s\right)\right)^2} \,  e^{-i\frac{\left(\k + \frac{m^2_{Q}}{E^2} \o \frac{\u}{2u^-}\right)^2}{2\o\O} \tan\left(\O\left(L-x^+_s\right)\right) } \, .
\end{align}
Finally, we notice that the Out-Out contribution in Eq.~\eqref{eq: initial squared amplitude} coincides with the vacuum spectrum, which is to be subtracted and thus drops out 
\begin{align}
     \o \frac{dI^{\text{MI}}}{d\o d^2\k}= \o \frac{dI^{\text{In-In}}}{d\o d^2\k}+\o \frac{dI^{\text{In-Out}}}{d\o d^2\k} \, .
\end{align}
We further redefine $dI=dI^{MI}$ and obtain the results presented above by numerically evaluating the remaining integrals.

\end{document}